\documentclass[namedreferences]{kluwer}

\def\that{\hat{\theta}}
\def\zhat{\theta^0}
\def\phat{\hat{\phi}}
\def\expt{{\sf E}}
\def\pa#1#2{\dfrac{\partial #1}{\partial #2}}
\def\papa#1#2#3{\dfrac{\partial^2 #1}{\partial #2 \partial #3}}
\def\dfrac#1#2{{\displaystyle\frac{#1}{#2}}}

\newtheorem{theorem}{Theorem}
\newtheorem{remark}{Remark}

\begin{document}
\begin{article}
\begin{opening}

\title{Application of the Information Criterion to the Estimation of 
  Galaxy Luminosity Function}
\author{Tsutomu T. \surname{Takeuchi}
  \thanks{Research Fellow of the Japan Society for the Promotion of Science}
  \email{E-mail: takeuchi@kusastro.kyoto-u.ac.jp}}
\institute{Department of Astronomy, Faculty of Science, Kyoto University, 
        606-8502, Japan}

\begin{abstract}
To determine the exact shape of the luminosity function  (LF) of galaxies 
is one of the central problems in galactic astronomy and observational 
cosmology.
The most popular method to estimate the LF is maximum likelihood, which is 
clearly understood with the concepts of the information theory.
In the field of information theory and statistical inference, great advance 
has been made by the discovery of Akaike's Information Criterion (AIC).
It enables us to perform a direct comparison among different types of models
with different numbers of parameters, and
provides us a common basis of the model adequacy.
In this paper we applied AIC to the determination of the shape 
of the LF.
We first treated the estimation using stepwise LF (\opencite{eep88}), 
and derived a formula to obtain the optimal bin number.
In addition, we studied the method to compare the goodness-of-fit of the 
parametric form \cite{sty79} with stepwise LF.
\end{abstract}
\end{opening}

\section{Introduction}

The luminosity function of galaxies (LF) is one of the fundamental
descriptions of the galaxy population (e.g. \opencite{bst88}).
It is also essential to interpret the galaxy number counts (e.g. 
\opencite{kk92}; \opencite{el97}) or to analyse galaxy clustering 
(e.g. \opencite{sw95}; \opencite{e96}).
Furthermore, the LF is a fundamental test for the theory of 
galaxy formation (e.g. \opencite{bcf96}).
Recently, the exact shape of the LF has been of particular interest, 
because it is one of the key issues to the ``faint blue galaxy problem'' 
of galaxy number counts (\opencite{kk92}; \opencite{el97}), and may be 
related to dwarf galaxy formation (e.g. \opencite{br92}; \opencite{bf96}
; \opencite{hp97}; \opencite{fb98}).

Instead of classical $V/V_{\rm m}$-estimator of the LF \cite{s68}, 
the maximum likelihood method, which is free of bias induced by density 
inhomogeneity, is popular among recent studies.
Various techniques have been proposed by expert astronomers 
(\opencite{ly71}; \opencite{m83}; \opencite{c86}; \opencite{cp93}).
The parametric method of \inlinecite{sty79} (STY) and
the stepwise maximum likelihood method by \inlinecite{eep88}(EEP) 
are the most popular among them.
And in addition, Lynden-Bell's method is quite sophisticated and requires
no assumption for the probability density function.
But in practice, we often need to smooth the obtained LF, because the 
discrete feature of the real data is not suitable for various studies.
When we perform smoothing or binning, it remains unclear, for example, 
that how many numbers of bins we should take.
Too wide bin leads to underestimation of the slope, and too narrow bin
makes the results unstable (e.g. \opencite{cp93}; \opencite{sw95}; 
\opencite{h97}).
It is also hard to estimate the relative goodness between different types 
of statistical models.
We should face this kind of problem, for example, when we try to examine 
whether the Schechter form \cite{sch76} provides an acceptable fit or 
not, using the likelihood ratio of the models (EEP).
In general, the more the number of free parameter is, the better the 
fitting becomes, and consequently, {\sl the likelihood function gets larger}.

In order to evaluate the goodness-of-fit of a certain model to the data, 
the concept of the information criterion is useful.
Since the middle of 1970's, vast advances have been made in the field of 
the statistical inference by the discovery of Akaike's Information Criterion 
(AIC: \opencite{akaike}).
The meaning of the AIC is clearly understood as an extention of
the maximum likelihood method, and closely related to the information 
entropy, especially to the `relative entropy' of two probability 
distributions.
The relative entropy has a property just like a distance in differential 
geometry, i.e. it is a distance between the two probability distributions.
Using AIC enables us to compare the goodness of a certain model with that of
another type directly.
For this fascinating property, AIC and its cousins are applied to 
the various fields of studies concerning statistical model selection.

In this paper, we make an attempt to apply AIC to the estimation problem 
of the LF.
In order to understand the Akaike's theory, some knowledges from information 
theory are required. Section 2 is devoted to this mathematical background 
concepts.
In section 3, first we see how AIC is applied to decide the step number of 
EEP method, and next how to judge the goodness of fit of the Schechter form 
estimated by STY method.
Our summary is presented in section 4.

\section{Akaike's information criterion}\label{section:aic}

In this section, we make an informal introduction of 
Akaike's theory.
We do not try to be mathematically rigorous, but
make an attempt to make it comprehensible.

\subsection{Kullback--Leibler information and information matrix}\label{bit}

First of all, we consider the `information entropy'.
Here we consider the (discrete) probability distribution 
$\{f_i\}_{i = 1, \cdots ,N}$.
The self information is defined as
\begin{equation}
  I (f_i) = - \ln f_i \; ,
\end{equation}
where $\ln f_i = \log_e f_i$.
Then, the expectation value of $I(f_i)$ is
\begin{equation}
  S \equiv {\textsf E}\left[I\right] = - \sum^N_{i = 1} f_i \ln f_i \; .
\end{equation}
This is called the information entropy.
When we have two different probability distributions $\{f_i\}_{i = 1, 
\cdots ,N}$ and $\{g_i\}_{i = 1, \cdots ,N}$, we can construct the 
following quantity:
\begin{equation}\label{ren}
  V(f, g) \equiv - \sum^N_{i = 1} f_i \ln \frac{f_i}{g_i} \; .
\end{equation}
This is called Kullback--Leibler information (\opencite{kl51}), 
or more comprehensively, relative entropy of $\{f_i\}_{i = 1, \cdots ,N}$ 
and $\{g_i\}_{i = 1, \cdots ,N}$.
For understanding the meaning of $V (f, g)$, we define $p_i = p_i 
(\theta_1, \cdots , \theta_K)$ ($i = 1, \cdots, N$), where 
$(\theta_1, \cdots , \theta_K)$ are the parameters on which the model depends, 
and let
\begin{equation}
  \left\{
    \begin{array}{@{\,}ll}
      f_i = p_i(\theta^0_1, \cdots , \theta^0_K) \; , \\
      \;  \\
      g_i = p_i(\theta^0_1 + {\rm d}\theta_1, \cdots , \theta^0_K + 
      {\rm d}\theta_K) \;.  
    \end{array}
  \right.
\end{equation}
Then, after some arithmetics, we have
\begin{eqnarray}
  V(f, g) = \frac{1}{2} \sum^K_{k=1}\sum^K_{l=1} I(\theta^0)_{kl} 
  {\rm d}\theta_k {\rm d}\theta_l \; ,
\end{eqnarray}
where
\begin{eqnarray}\label{infomatrix}
  I_{kl}(\theta^0) = \sum^N_{i=1}
  p_i \left(\pa{\ln p_i}{\theta_k} \pa{\ln p_i}{\theta_l} 
  \right)_{\theta = \theta^0}\; .
\end{eqnarray}
The subscript $\theta = \theta^0$ means $\theta_i = \theta^0_i$ for
$i = 1, \cdots , N$.
The matrix $I_{kl}$ is called Fisher's information matrix.
If we regard $I_{kl}$ as a ``metric'' of the parameter space of $\theta_k$
and $\theta_l$, we can treat $V(f, g)$ as the distance between 
the two probability distributions
$\{f_i\}_{i = 1, \cdots ,N}$ and $\{g_i\}_{i = 1, \cdots ,N}$,
just as we do in differential geometry.

\subsection{Maximum likelihood estimation in the context of information theory}

We, here, will see the maximum likelihood method considering relative 
entropy $V(f, g)$.
Maximum likelihood is the method in order to estimate the optimal model 
for {\sl a given set of data}.
Let ${x_1, \cdots, x_N}$ as a realization of random variable
$X$ which obeys the unknown probability distribution $f(X)$.
We define a model probability distribution   
$g(x;\theta_1, \cdots, \theta_K)$, which depends on $K$ 
parameters $\{\theta_k\}_{k = 1, \cdots, K}$.
Then, the likelihood function $\cal{L}$ is defined by
\begin{eqnarray}
  {\cal L} \equiv
  {\cal L} (\theta_1, \cdots, \theta_K | x_1, \cdots, x_n) 
  = \prod^N_{n=1} g(x_n; \theta)\; ,
\end{eqnarray}
and the logarithmic likelihood is
\begin{equation}
   \ln {\cal L}\equiv \sum^N_{n=1} \ln g(x_n; \theta )\; .
\end{equation}
Performing maximum likelihood estimation is to find the parameter set
$\theta = \theta_1, \cdots, \theta_K$ which defines the most suitable 
model $g(x;\theta)$ for unknown {\sl true} probability distribution 
$f(X)$.
If we know the true $f(X)$, using relative entropy eq.(\ref{ren}), 
we have 
\begin{eqnarray}\label{distance}
  V(f, g) &=&  
  \sum^N_{n=1} f(x_n) \ln \dfrac{f(x_n)}{g(x_n;\theta)} 
  = {\textsf E}\left[\,\ln \dfrac{f(X)}{g(X; \theta)}\right] \; ,\nonumber \\
  && \nonumber \\
  &=&  
  {\textsf E}\left[\,\ln f(X)\right] - {\textsf E}\left[\ln g(X;\theta)\right] 
  \; .
\end{eqnarray}
But actually we do not know $f(X)$, we cannot obtain ${\textsf 
E\left[\;\cdot\;\right]}$.
Therefore we use the law of large numbers.

\begin{theorem}\label{lln}
{\bf (The law of large numbers)}
{\sl
Let $X_1, \cdots, X_N$ independent random variables, which have the 
same mean value
\begin{equation}
  {\expt}\left[X_1\right] = \cdots = {\expt}\left[X_N\right] = m \; ,
\end{equation}
and ${}^\exists \sigma^2 > 0$, such that
\begin{equation}
  {\expt}\left[(X_i - m_i)\right] \le \sigma^2  \quad {\rm for} \quad 
  i = 1, \cdots , N\; .
\end{equation}
Then for a random variable
\begin{equation}
  X \equiv \frac{1}{N} \sum^N_{i=1} X_i \; ,
\end{equation}
we have 
\begin{equation}
  {\expt}\left[X\right] = m \; ,
\end{equation}
as $N \rightarrow \infty$.
}
\end{theorem}
Proof of this famous theorem is seen in many textbooks on probability theory,
so we omit it (see e.g. \opencite{shi96}).

Hence, if the size of the dataset $N$ is sufficiently large,
\begin{equation}
  \frac{1}{N}\ln {\cal L} =  \frac{1}{N}\sum^N_{i=1} \ln g(x_n;\theta) 
  \simeq {\expt} \left[\ln g(X;\theta)\right] \;.
\end{equation}
It leads to 
\begin{equation}\label{eq:klofml}
  V(f, g) = {\expt} \left[\ln f(X)\right] - 
  \frac{1}{N}\sum^N_{i=1} \ln g(x_n;\theta) \; .
\end{equation}
The last term can be obtained without knowing true $f(X)$.
Since ${\expt} \left[\ln f(X)\right]$ is independent of $\theta = 
\theta_1, \cdots , \theta_K$, 
what to be maximized is $\ln {\cal L} =  \sum^N_{i=1} 
\ln g(x_n;\theta)$ .
Consequently, we obtain the likelihood equation: 
\begin{equation}\label{loglik}
  \left.\pa{\;}{\theta_k}\ln {\cal L} \right|_{\theta = \hat{\theta}}
  =  \pa{\;}{\theta_k}\left\{ \sum^N_{n=1} \ln g(x_n; \theta) 
  \right\}_{\theta = \hat{\theta}} = 0 
\end{equation}
for $k = 1, \cdots, K$, where we defined $\hat{\theta}$ to satisfy
\begin{equation}
  \max_{\theta \in \Theta} (\ln {\cal L}) 
 = \ln {\cal L} (\hat{\theta}| x_1, \cdots, x_n)\; .
\end{equation}
Here, $\Theta$ denotes the family of parameters $\theta$.
The solution $\that$ of the eq. (\ref{loglik}) is called 
the maximum likelihood estimator.

We, next, consider the error of the logarithmic likelihood equation 
(\ref{loglik}):
\begin{equation}\label{error}
  \Delta (\that) \equiv \left. \expt\left[\ln g(X; \theta) \right] 
  \right|_{\theta = \that}
  - \frac{1}{N}\sum^N_{n=1} \ln g (x_n; \theta) |_{\theta = \that}\; .
\end{equation}
Let $\zhat = \zhat_1, \cdots, \zhat_K \in \Theta$ the estimator which maximize 
$\expt\left[\ln g(X;\theta)\right]$.
Taylor expansion and theorem \ref{lln} lead the well-known important result:
\begin{eqnarray}
  \Delta(\that) 
  &\simeq&
  \left. \sum^N_{k=1}\sum^{N}_{l=1}
    (\that_k - \zhat_k) \expt\left[\papa{\;}{\theta_k}{\theta_l} 
      \ln g(X;\theta)\right]
  \right|_{\theta = \zhat} (\that_l - \zhat_l)  \nonumber \\
  && \nonumber \\
  &\simeq&
  - \sum^N_{k=1}\sum^{N}_{l=1}(\that_k - \zhat_k) I_{kl} (\that_l - \zhat_l)\;.
\end{eqnarray}
Again, $I_{kl}$ is the information matrix.
Thus, we can evaluate the error of likelihood function using $I_{kl}$ 
(e.g. \opencite{soa99}).

\subsection{Akaike's Information Criterion (AIC)}

In this subsection, we discuss Akaike's information criterion (AIC).
To the first order of $\theta$, likelihood equation (\ref{loglik}) 
is expressed as
\begin{eqnarray}\label{fole}
  &&\sum^N_{n=1} \left\{ \pa{\;}{\theta_k}\ln g(x_n;\theta)
  \right\}_{\theta = \zhat}
  + \sum^N_{n=1}\sum^K_{l=1} \left\{ \papa{\;}{\theta_k}{\theta_l}
    \ln g(x_n; \theta) \right\}_{\theta = \zhat}(\that_l - \zhat_l) 
  \nonumber \\
  &&= 0 \; .
\end{eqnarray}
In order to evaluate this equation, we use the multivariate central limit
theorem.
\begin{theorem}\label{mclt}
{\bf (Multivariate central limit theorem)}
{\sl 
Let $a^k_n$ $(n = 1, \cdots, N)$ a set of sample values of a vector random 
variable $A^k$, which have a mean 
\begin{eqnarray}
  \expt_A\left[A^k\right] = m^k\; , 
\end{eqnarray}
($m^k < \infty$) and a dispersion
\begin{eqnarray}
  \expt_A \left[(A^k - m^k)(A^l - m^l)\right] = \sigma^{kl}
\end{eqnarray}
for $(k = 1, \cdots, K)$.
We define $Z^k$ as a vector random variable whose sample value $z^k$ is 
\begin{eqnarray}
  z^k \equiv \sqrt{N} \left\{ \frac{1}{N} \sum^N_{n=1} 
    a^k_n - m^k \right\} \; .
\end{eqnarray}
Then, as $N \rightarrow \infty$, $Z^k$ becomes a Gaussian vector random 
variable with
\begin{eqnarray}
  \expt_Z\left[Z^k\right] = 0\; , \quad
  \expt_Z\left[Z^kZ^l\right] = \sigma^{kl}\; .
\end{eqnarray}
}
\end{theorem}
The proof of this theorem is also incorporated in many textbooks (e.g. 
Shiryaev 1996).

By definition of $\zhat$, we have
\begin{eqnarray}
  \left.\expt\left[\pa{\;}{\theta_k} \ln g(X; \theta)\right]
  \right|_{\theta = \zhat} 
  = \left\{\pa{\;}{\theta_k} \expt\left[\ln g(X; \theta)\right] 
  \right\}_{\theta = \zhat} \equiv {\cal M}^k = 0 \; . 
\end{eqnarray}
Therefore, with eq. (\ref{infomatrix})
\begin{eqnarray}
  &&  \left.\expt\left[\left(\pa{\;}{\theta_k} \ln g(X; \theta)-{\cal M}^k 
      \right)
      \left( \pa{\;}{\theta_l} \ln g(X; \theta)-{\cal M}^l \right) \right]
  \right|_{\theta = \zhat} \nonumber \\
  &&= \left.\expt\left[\left(\pa{\;}{\theta_k} \ln g(X; \theta)\right)
      \left( \pa{\;}{\theta_l} \ln g(X; \theta)\right) \right]
  \right|_{\theta = \zhat} \nonumber \\
  &&= I_{kl} \; .
\end{eqnarray}
Using the result of theorem \ref{mclt}, we have a vector random variable 
$Z^k$ which is a Gaussian with
\begin{eqnarray}
  \expt_Z [Z^k] = 0 \; , \; \expt_Z[Z^kZ^l] = I_{kl}\; ,
\end{eqnarray}
and whose sample value $z^k$ is 
\begin{eqnarray}\label{cesaro}
  z^k = \frac{1}{\sqrt{N}} \sum^N_{n=1}\left\{ \pa{\;}{\theta_k} 
    \ln g(x_n; \theta) \right\}_{\theta = \zhat} \; .
\end{eqnarray}
We, then, apply theorem \ref{lln} to the second term of the 
eq.(\ref{fole}), it follows that
\begin{eqnarray}\label{applclt}
\frac{1}{N} \sum^N_{n=1}\left\{ \papa{\;}{\theta_k}{\theta_l} 
  \ln g(x_n; \theta) \right\}_{\theta = \zhat} 
  &\simeq& \left.\expt\left[ \papa{\;}{\theta_k}{\theta_l}\ln g(X; \theta) 
    \right]\right|_{\theta = \zhat} \nonumber \\
  &=& - I_{kl}\, .
\end{eqnarray}
Substituting the eqs. (\ref{cesaro}) and (\ref{applclt}) into eq. (\ref{fole}) 
gives
\begin{eqnarray}
  z^k = \sqrt{N} \sum^K_{l=1} I_{kl} (\that_l - \zhat_l)  \; .
\end{eqnarray}
If $I_{kl}$ is non-singular (We can always expect it by 
reparametrization), the inverse matrix $I^{-1}_{kl}$ is defined as
\begin{eqnarray}
  \sum^K_{k=1} I_{uk}I^{-1}_{kv} = \delta_{uv}
\end{eqnarray}
where $\delta_{uv}$ is a Kronecker's delta, and it leads to
\begin{eqnarray}
  (\that_k - \zhat_k) = \frac{1}{\sqrt{N}} \sum^K_{l=1} I^{-1}_{kl}z^l \; .
\end{eqnarray}
Hence, we can write
\begin{eqnarray}
  \Delta(\that) = -\sum^K_{k=1}\sum^K_{l=1}(\that_k - \zhat_k)I_{kl}
  (\that_l - \zhat_l) 
  = -\frac{1}{N}\sum^K_{u=1}\sum^K_{v=1}I^{-1}_{uv}z^uz^v \; .
\end{eqnarray}
Now, again we use theorem \ref{mclt}, and eventually we obtain, 
\begin{eqnarray}
  \Delta(\that) 
  &\simeq&  
  -\frac{1}{N}\sum^K_{u=1}\sum^K_{v=1}I^{-1}_{uv}
  \expt_Z [Z^uZ^v] =  
  -\frac{1}{N}\sum^K_{u=1}\sum^K_{v=1}I^{-1}_{uv}I_{uv} \nonumber \\
  &=&  -\frac{K}{N}\; .
\end{eqnarray}
Thus, we reach final important result:
\begin{eqnarray}
  \left.\expt[\ln g(X; \theta)] \right|_{\theta = \that} 
  &\simeq&  
  \frac{1}{N} \left.\sum^N_{n=1} \ln g(x_n; \theta) \right|_{\theta = \that}
  - \frac{K}{N} \nonumber \\
  &&\nonumber \\
  &=&  
  \frac{1}{N} \left(\ln {\cal L}(\that | x_1, \cdots, x_n) - K \right)\; .
\end{eqnarray}
Since, as eq. (\ref{distance}) shows, the distance between the true 
probability distribution $f(X)$ and model $g(X;\theta)$ is evaluated by
$\expt[\ln g(X;\theta)]$, we have to choose $\that \in \Theta$ to be 
maximize the quantity $(\ln {\cal L} - K)$ for given $K$.
Though large $K$ provides us large $\ln {\cal L}$, it also reduces 
$(\ln {\cal L} - K)$ at the same time, therefore the optimal number of 
parameters $K$ has to be taken so that it maximize
$(\ln {\cal L}(\that) - K)$. 
For historical reason (e.g. \opencite{r65}), the value 
$-2(\ln {\cal L}(\that)-K)$ has been considered (see section 
\ref{section:discussion}).
It is the Akaike's information criterion (AIC).
We, now, see that the optimal model is the one which minimizes AIC.

\section{Application of AIC to LF estimation}

\subsection{The stepwise maximum likelihood method and optimal step number}

We apply AIC to the estimation of the shape of the LF.
First we consider the stepwise maximum likelihood method introduced by EEP.
Throughout this paper, for simplicity, we do not deal with some issues 
which turn out to be important when we treat the real redshift data, 
e.g. errors in magnitude measurement, surface brightness incompleteness, etc.
These practical problems are summarized and dealt with in e.g. 
\inlinecite{l96}.

The EEP method uses the form of the LF
\begin{eqnarray}
  \phi (M) = \sum^K_{k=1} \phi_k  W(M_k - M) 
\end{eqnarray}
where $M$ is the absolute magnitude of a galaxy, 
which is obtained by
\begin{eqnarray}
  M = m - 25 - 5\log d_L(z) - k(z) \; .
\end{eqnarray}
Here $\log \equiv \log_{10}$, $m$ : the apparent magnitude, $d_L (z)$ [Mpc] : the luminosity distance
corresponding to redshift $z$, and $k(z)$ is the $K$-correction.
The window function  $W(M_l - M)$ is defined by
\begin{eqnarray}
  W(M_l - M) \equiv 
  \left\{
    \begin{array}{@{\,}ll}
      1 & M_l - \Delta M/2 \leq M \leq M_l + \Delta M/2\;, \\
      ~&~  \\
      0 & {\rm otherwise} \; .
    \end{array}
  \right.
\end{eqnarray}
In terms of information theory, the stepwise LF is regarded as a histogram 
model, a kind of discrete probability distribution models.
The parameters of the model $\theta \in \Theta$ is in this case $\theta_k 
= \phi_k$ ($k = 1, \cdots , K$) themselves.
Given the galaxy redshift survey data of size $N$, we set
\begin{eqnarray}\label{deltam}
 &&\Delta M = \dfrac{M_{\rm upper} - M_{\rm lower}}{K - 1} \; , \\
 && \nonumber \\
 && M_{\rm upper} \equiv \max_{i = 1, \cdots , N} \{M_i\} \; , \; 
    M_{\rm lower} \equiv \min_{i = 1, \cdots , N} \{M_i\} \; .\nonumber
\end{eqnarray}
The denominator of the eq. (\ref{deltam}) is $K-1$, though the number of 
bins are $K$, because $M_k$ is evaluated at the center of $k$-th bin, and
therefore the magnitude range becomes $M_{\rm lower} - \Delta M/2 \sim 
M_{\rm upper} + \Delta M/2$.
A certain constraint is adopted to $\{\phi_k \}_{k=1, \cdots , K}$ in usual 
manner.
However, since the normalization does not used in the determination of the 
shape of the LF (otherwise a Lagrange multiplier $\lambda$ would appear in 
the denominator of the eq. (\ref{solution})), and neither adopted in the 
procedure of the following STY method, we do not set any constraint in 
our formulation.
According to EEP, the likelihood function is
\begin{eqnarray}
  &&{\cal L}(\{ \phi_k \}_{k=1, \cdots ,K}| \{ M_i \}_{i=1, \cdots ,N})
  \nonumber \\
  &&= \prod^N_{i=1}
  \dfrac{\sum^K_{l=1}W(M_l - M_i)\phi_l}{\sum^K_{l=1}\phi_l 
  H(M_{\rm lim}(z_i) - M_l) \Delta M}\;,
\end{eqnarray}
\begin{eqnarray}
  &&  H (M_{\rm lim}(z_i) - M_l) \nonumber \\
  && \equiv  \left\{
    \begin{array}{@{\,}ll}
     1 & M_{\rm lim}(z_i) - \Delta M/2 > M_l \\
     ~ & ~ \\
     \dfrac{M_{\rm lim}(z_i) - M_l}{\Delta M} + \dfrac{1}{2} &
     M_{\rm lim}(z_i) - \Delta M/2 \leq \\
     ~ & \quad M_l < M_{\rm lim}(z_i) + \Delta M/2 \\
     0 & M_{\rm lim}(z_i) + \Delta M/2 \leq M_l
    \end{array} 
  \right. 
\end{eqnarray}
where $M_{\rm lim}(z_i)$ is the absolute magnitude corresponding to the 
survey limit $m_{\rm lim}$ at redshift $z_i$.
This likelihood 
function ${\cal L}$ clearly depends on the bin width $\Delta M$, and 
consequently, its likelihood ratio to other model depends on $\Delta M$.
This has been regarded as an ``artificial effect'' to be eliminated by 
certain procedures, but it is not true because the choice of $\Delta M$ is 
in this case the selection of histogram model itself.

The logarithmic likelihood is expressed as
\begin{eqnarray}
  &&\ln {\cal L} = 
  \sum^N_{i=1} \left[
  \sum^K_{l=1}W(M_l - M_i) \ln \phi_l \right.\nonumber \\
  &&- 
  \left.\ln \left\{ \sum^K_{l=1}\phi_l H(M_{\rm lim}(z_i) - M_l)\Delta M
  \right\} 
  \right] \; .
\end{eqnarray}
Hence, likelihood equation becomes
\begin{eqnarray}
  &&\pa{\ln {\cal L}}{\phi_k} =
  \sum^N_{i=1} \frac{W(M_k - M_i)}{\phi_k} \nonumber \\
  &&-   \sum^N_{i=1} 
  \dfrac{H(M_{\rm lim}(z_i) - M_k) 
  \Delta M}{\sum^K_{l=1}\phi_l H(M_{\rm lim}(z_i) - M_l) \Delta M} = 0
\end{eqnarray}
and it reduces to 
\begin{eqnarray}\label{solution}
  \phi_k \Delta M = 
  \dfrac{\sum^N_{i=1} 
  W(M_k - M_i)}{\sum^N_{i=1} 
  \dfrac{H(M_{\rm lim}(z_i) - M_k)}{\sum^K_{l=1}
  \phi_l H(M_{\rm lim}(z_i) - M_l) \Delta M}}\; .
\end{eqnarray}
This equation can be solved by iteration, and we obtain the maximum likelihood
estimator $\phat = \{\phat_k\}_{k=1, \cdots, K}$.
Thus, 
\begin{eqnarray}\label{aic}
  {\rm AIC_{EEP}} = -2( \ln {\cal L} |_{\phi = \phat} - K) \; .
\end{eqnarray}
Therefore, the step number $K$ should be taken so that it minimizes the eq.
(\ref{aic}).
\begin{remark} ~
{\sl
The obtained number $K$ may not stand for the number of physical parameters,
viz. when we get a certain $K$, it does not mean we need $K$ physical 
quantities for explanation.
The obtained stepwise LF is the one which best reflects the property of the 
underlying data population.
}
\end{remark}

\subsection{Comparison of Schechter form with stepwise LF}

We, here, consider how to compare the goodness of fit of the Schechter 
form to that of EEP stepwise LF.
We set the LF as
\begin{eqnarray}\label{schlf}
  \phi(M) = 0.4\ln 10 \; \phi^\ast \left( 10^{0.4(M_\ast - M)} 
  \right)^{1+\alpha} \exp \left( -10^{0.4(M_\ast - M)} \right) \; .
\end{eqnarray}
The likelihood function is therefore
\begin{eqnarray}
  &&{\cal L}(\{\alpha, M_\ast \} |\{ M_i \}_{i = 1, \cdots, N}) \nonumber \\
  &&= \prod^N_{i=1} 
  \dfrac{\displaystyle \mathstrut 
  \left( 10^{0.4(M_\ast - M_i)} \right)^{1+\alpha} 
  \exp \left( -10^{0.4(M_\ast - M_i)} \right)}{\displaystyle \mathstrut
  \int^{M_{\rm lim}(z_i)}_{-\infty} 
  \left( 10^{0.4(M_\ast - M)} \right)^{1+\alpha} 
  \exp \left( -10^{0.4(M_\ast - M)} \right) {\rm d}M} \; .
\end{eqnarray}
Thus the logarithmic likelihood becomes
\begin{eqnarray}
  \ln{\cal L} 
   &=&  
  0.4\ln{10} \, (1+\alpha) \sum^N_{i=1} (M_\ast - M_i) 
  - \sum^N_{i=1} 10^{0.4(M_\ast - M_i)} \nonumber \\
  &&- \sum^N_{i=1} \ln \int^{M_{\rm lim}(z_i)}_{-\infty} \!\!
  \left( 10^{0.4(M_\ast - M)} \right)^{1+\alpha} 
  \exp \left( -10^{0.4(M_\ast - M)} \right) {\rm d}M \nonumber \\
  &\; & \nonumber \\
   &=&  
  0.4\ln{10} \, (1+\alpha) \sum^N_{i=1} (M_\ast - M_i) \nonumber \\
  &&- \sum^N_{i=1} 10^{0.4(M_\ast - M_i)}
  - 0.4\ln{10} \sum^N_{i=1} \ln \Gamma (1+\alpha , y_i) \; , 
\end{eqnarray}
where
\begin{eqnarray}
  && y_i \equiv 10^{0.4 (M_\ast-M_{\rm lim}(z_i))}\; , \\
  && \Gamma (z, p) \equiv \int^\infty_p t^{z-1}e^{-t}{\rm d}t \; .
  \label{igamma}
\end{eqnarray}
Equation (\ref{igamma}) is known as Legendre's incomplete gamma function.

Since the parameters to be estimated are $\alpha$ and $M_\ast$ in the 
STY method, we have the set of likelihood equations:
\begin{eqnarray}
  &&\pa{\ln{\cal L}}{\alpha} \nonumber \\
  &&=  
  0.4\ln{10} \left\{ \sum^N_{i=1} (M_\ast - M_i)
    - \sum^N_{i=1} \dfrac{\int^\infty_{y_i} \ln t \; t^\alpha e^{-t}
  {\rm d}t}{\Gamma (1+\alpha, y_i)}\right\} = 0 \; , \label{leqalpha}\\
  &&\pa{\ln{\cal L}}{M_\ast} =  
  0.4\ln{10} (1+\alpha)N  \nonumber \\
  &&- 0.4\ln{10} \sum^N_{i=1} {10}^{0.4(M_\ast - M_i)} 
  - 0.4\ln{10} \sum^N_{i=1} \dfrac{{y_i}^\alpha e^{-y_i}}{\Gamma 
    (1+\alpha, y_i)} \pa{y_i}{M_\ast} \nonumber \\
  &&= 0.4\ln{10}\nonumber \\
  &&\times \left\{ (1+\alpha) N 
    - \sum^N_{i=1} 10^{0.4(M_\ast - M_i)} 
    - 0.4\ln{10}\dfrac{{y_i}^{1+\alpha}e^{-y_i}}{\Gamma (1+\alpha, y_i)} 
  \right\}\nonumber \\
  &&= 0 \; \label{leqmag}.
\end{eqnarray}
By solving the eqs. (\ref{leqalpha}) and (\ref{leqmag}), we obtain the 
maximum likelihood estimators $\hat{\alpha}$ and $\hat{M_\ast}$.
We have two free parameters in the STY method, and we consequently have
the AIC of STY method
\begin{eqnarray}
{\rm AIC_{STY}} = -2(\ln {\cal L}_{\rm STY} |_{\theta = \that} - 2)\;, 
\end{eqnarray}
where the subscript $\theta = \that$ represents that $\alpha =\hat{\alpha}$ and
$M_\ast = \hat{M_\ast}$.

The relative goodness of fit of the Schechter LF compared with the stepwise 
one is evaluated by 
\begin{eqnarray}\label{eq:deltaaic}
  \Delta{\rm AIC} 
  &\equiv&  {\rm AIC}_{\rm STY} - {\rm AIC}_{\rm EEP} \nonumber \\
  &=&  -2(\ln {\cal L}_{\rm STY} |_{\theta = \that} 
  - \ln {\cal L}_{\rm EEP} |_{\phi = \hat{\phi}} + K - 2 ) \; .
\end{eqnarray}

\section{Summary and discussion}\label{section:discussion}

In in the previous sections we introduced the Akaike's information 
criterion (AIC) to the maximum likelihood estimation of galaxy luminosity 
function (LF).
The AIC is closely related to the ``distance'' between two probability 
distributions which becomes clear by using Fisher's information matrix.
It is expressed as 
\begin{eqnarray}
  {\rm AIC} = -2(\ln {\cal L}(\that)-K)\; ,\nonumber
\end{eqnarray}
where $\cal L$ is a likelihood function, $\that$ is a set of maximum 
likelihood estimators, and $K$ is the number of free parameters of the 
assumed model.
Since the concept of the information criterion seems unfamiliar to the 
astronomical community, we discuss its meaning and practical use in
this section.

What we must stress is that the difference between the $\chi^2$-type 
goodness-of-fit quantities and AIC.
The statistic usually used to evaluate uncertainty of the LF
estimation is the $\chi^2$, which was extensively discussed by EEP.
This error evaluation is based on the well-known fact that the logarithmic 
likelinood ratio, $-2 \ln ({\cal L}(\theta)/{\cal L}(\zhat))$, 
is asymptotically distributed as $\chi^2$ distribution (e.g. \opencite{r65}; 
\opencite{soa99}).
The likelihood ratio is regarded as a random variable, and discussed with 
respect to its confidence level.
In order to estimate its distribution, EEP performed a Monte Carlo 
simulations and confirmed their error estimation.
On the other hand, as we discussed in section \ref{section:aic}, the 
information criterion is a value obtained as a result of limit theorems, 
and is not regarded as a random variable.
In statistical model selection, we often use such kind of goodness-of-fit 
index.
The class of information criteria including AIC was invented along with 
such concepts.
Thus, though the AIC is related to the $\chi^2$ statistic, likelihood ratio,
its value is not discussed with a confidence level (see \opencite{s86} for
details).

Then, how should we treat the AIC value for practical use?
In the case of stepwise LF model (EEP), $K$ is a number of steps of the LF. 
The AIC is 
\begin{eqnarray}
  {\rm AIC_{EEP}} = -2( \ln {\cal L} |_{\phi = \phat} - K) \; , \nonumber
\end{eqnarray}
where $\phi = \phi_1, \cdots, \phi_K$ is the step heights.
We should compare the ${\rm AIC_{EEP}}$ and choose the number $K$ which 
minimizes ${\rm AIC_{EEP}}$.
The larger the number of parameter is, the larger datasize $N$ is required, 
because each parameter estimation procedure carries its own error.
In case $K$ is considerably larger in comparison with the data size $N$, we 
cannot use the limit theorems of probability theory like theorem \ref{lln},
since the results would no longer hold.
Thus, number of parameters should be taken as 
\begin{equation}
  K \lsim 2\sqrt{N}
\end{equation}
and at most
\begin{equation}
  K < \frac{N}{2} \; .
\end{equation}
Bin number $K$ comparable to $N$ is meaninglessly large 
because such a fine binning yields horribly numerous empty bins.
In such case, the AIC diverges by its definition, which means that 
the bin number is not a good choice.
Sometimes, AIC becomes smaller and smaller as $K$ is larger, and does not 
take minimum value.
Then the assumed form of the model is significantly wrong, so that 
$V(f, g)$ is very large.

We next consider the comparison between the stepwise form and parametric 
models.
For the STY method, free parameters are $\alpha$ and $M_\ast$ in the 
Schechter function (eq.(\ref{schlf})).
In order to compare the goodness of fit of EEP and STY models, we can use
the difference of the AIC
\begin{eqnarray}
  \Delta{\rm AIC} 
  &\equiv&   {\rm AIC}_{\rm STY} - {\rm AIC}_{\rm EEP} 
  \nonumber \\
  &=&  -2(\ln {\cal L}_{\rm STY} |_{\theta = \that} 
  - \ln {\cal L}_{\rm EEP} |_{\phi = \hat{\phi}} + K - 2 ) \; , \nonumber
\end{eqnarray}
where the $\theta = \that$ means that $\alpha =\hat{\alpha}$ and
$M_\ast = \hat{M_\ast}$.
As we mentioned above, the evaluation of AIC is essentially different from
that of $\chi^2$ statistic.
Originally, EEP regarded their stepwise LF as a {\sl true} probability 
density function and derived the likelihood ratio.
They mentioned that the likelihood ratio obtained from stepwise and 
Schechter LF was very large and negative.
This fact is a natural consequence, because, from the aspect of 
(relative) Kullback--Leibler distance, stepwise LF has much larger 
parameters compared with parametric Schechter LF and therefore the 
goodness-of-fit is much better.
Information criterion is suitable for such problem.

Again we note that it is the difference of AIC values that matters and 
not the absolute values themselves.
This is because we would never know the ture distribution from a finite 
size data, and we used the sample expectation instead of the expectation
value based on true distribution (see eq. (\ref{eq:klofml})).
In our point of view, the stepwise LF is an estimates derived along with 
one of the family of statistical models in this paper.
Thus the difference of AICs, $\Delta{\rm AIC}$ can be used for our purpose.
Then, what should we regard as the ``scale'' of $\Delta{\rm AIC}\;$? 
The order of the variation of the AIC is that of the number of parameters, 
$K$, which is an integer.
Thus, when we have the difference of AICs
\begin{eqnarray}
  \Delta{\rm AIC} \gsim 1\; ,
\end{eqnarray}
the two distributions are significantly different.
Generally, $\Delta{\rm AIC} = {\rm AIC}_{\rm STY} - {\rm AIC}_{\rm EEP}$
is larger than unity, therefore we judge that the goodness-of-fit is not
sufficient, and other functional form can instead be used.
But acutually, for the optical galaxy LF, we are interested in comparison
between the goodness-of-fit of Schechter form and other forms.
The functional form choice is also an interesting issue in the estimaiton 
of the {\sl IRAS}~ galaxy LF, which is known to be significantly differnt 
from Schechter form (e.g. \opencite{s90}).
In such case AIC works as a powerful tool for model selection of the 
fitting functions.

\begin{acknowledgements}
We offer our gratitude to Drs. Hiroyuki Hirashita and Kohji Yoshikawa 
who have made critical reading of the original version of the paper.
We also thank Drs. Toru Yamada, Takashi Ichikawa, Hajime Sugai, and 
Takehiko Wada for useful discussions and comments.
We acknowledge the Research Fellowships
of the Japan Society for the Promotion of Science for Young
Scientists.
\end{acknowledgements}

\end{article}
\end{document}